\documentclass[journal]{IEEEtran}

\usepackage{cite}

\ifCLASSINFOpdf
\else
\fi

\usepackage{amsmath}
\usepackage{multirow}
\usepackage{subfig}
\usepackage{graphicx}
\usepackage{adjustbox}
\usepackage{caption}
\usepackage{xcolor}

\begin{document}

\title{Robust and Interpretable Temporal Convolution Network for Event Detection in Lung Sound Recordings}



\author{Tharindu~Fernando,~\IEEEmembership{Member,~IEEE,}
        Sridha~Sridharan,~\IEEEmembership{Life Senior Member,~IEEE,}
        Simon~Denman,~\IEEEmembership{Member,~IEEE,}
        Houman~Ghaemmaghami,
        ~and~ Clinton~Fookes,~\IEEEmembership{Senior Member,~IEEE.}
     
\IEEEcompsocitemizethanks{\IEEEcompsocthanksitem T. Fernando, S. Sridharan, S.Denman and C. Fookes are with The Signal Processing, Artificial Intelligence and Vision Technologies (SAIVT), Queensland University of Technology, Australia.\protect\\ H. Ghaemmaghami is with M3DICINE Pty Ltd, Brisbane, Australia. }}

\markboth{Journal of \LaTeX\ Class Files,~Vol.~14, No.~8, August~2015}%
{Shell \MakeLowercase{\textit{et al.}}: Bare Demo of IEEEtran.cls for IEEE Transactions on Magnetics Journals}

\IEEEtitleabstractindextext{%
\begin{abstract}
\textit{Objective}: This paper proposes a novel framework for lung sound event detection, segmenting continuous lung sound recordings into discrete events and performing recognition on each event. Exploiting the lightweight nature of Temporal Convolution Networks (TCNs) and their superior results compared to their recurrent counterparts, we propose a lightweight, yet robust, and completely interpretable framework for lung sound event detection.
\textit{Methods}: We propose the use of a multi-branch TCN architecture and exploit a novel fusion strategy to combine the resultant features from these branches. This not only allows the network to retain the most salient information across different temporal granularities and disregards irrelevant information, but also allows our network to process recordings of arbitrary length.
\textit{Results}: The proposed method is evaluated on multiple public and in-house benchmarks of irregular and noisy recordings of the respiratory auscultation process for the identification of numerous auscultation events including inhalation, exhalation, crackles, wheeze, stridor, and rhonchi. We exceed the state-of-the-art results in all evaluations. Furthermore, we empirically analyse the effect of the proposed multi-branch TCN architecture and the feature fusion strategy and provide quantitative and qualitative evaluations to illustrate their efficiency. Moreover, we provide an end-to-end model interpretation pipeline that interprets the operations of all the components of the proposed framework.
\textit{Conclusion}: Our analysis of different feature fusion strategies shows that the proposed feature concatenation method leads to better suppression of non-informative features, which drastically reduces the classifier overhead resulting in a robust lightweight network.
\textit{Significance}: Lung sound event detection is a primary diagnostic step for numerous respiratory diseases. The proposed method provides a cost-effective and efficient alternative to exhaustive manual segmentation, and provides more accurate segmentation than existing methods. As such, it can improve the performance of subsequent processes such as diagnosis. The lightweight nature of our model allows it to be deployed in end-user devices such as smartphones, and it has the ability to generate predictions in real-time. The end-to-end model interpretability helps to provide the required trust in the system for use in clinical settings.
\end{abstract}

}

\maketitle

\IEEEdisplaynontitleabstractindextext

\IEEEpeerreviewmaketitle

\section{Introduction}

\IEEEPARstart{R}{espiratory} auscultation \cite{pham2021cnn, hsu2021benchmarking, raj2020nonlinear}, where lung sounds are captured using a stethoscope placed on a patient's chest or back, is a primary diagnostic step for respiratory diseases. In addition to inferring breathing cycles (inhalation and exhalation), numerous works \cite{hsu2021benchmarking, raj2020nonlinear, messner2018crackle} have shown how respiratory auscultation can be utilised for disease diagnosis. 

In \cite{hsiao2020breathing} the authors illustrate how lung sounds can be broadly categorised into two classes: normal breathing sounds; and adventitious breathing sounds. Adventitious breathing sounds usually indicate the presence of pulmonary diseases, which can be either continuous (ex: wheeze, stridor, rhonchi) or discontinuous (ex: crackles). However, in addition to simply identifying the presence of adventitious sound, the temporal location of the adventitious sound within the patient's respiratory cycle carries clinical significance. For instance, \cite{who} shows that fine-crackles occurring mid-to-late within the inhalation stage are indicative of interstitial lung fibrosis, congestive heart failure and pneumonia; while coarse-crackles during the early stages of inhalation and throughout exhalation are linked to chronic bronchitis. 

\begin{figure}[tbp]
    \centering
    \includegraphics[width=\linewidth]{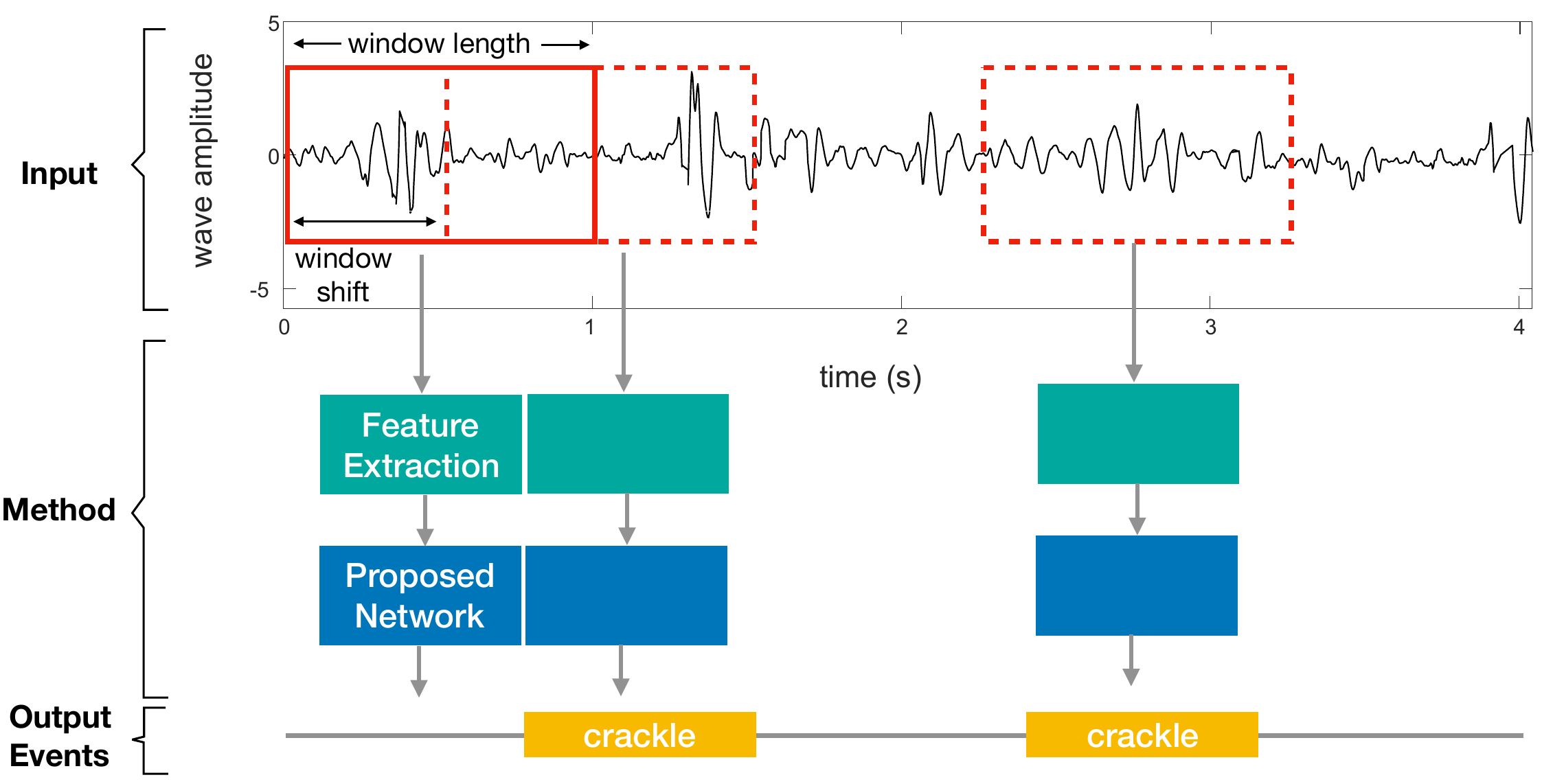}
    \caption{Proposed Event Detection Framework: We apply a sliding window of size 1 sec with 50\% overlap from the input lung sound recording. Extracted windows are passed through the feature extractor and then through the proposed network which outputs the event category for the window.}
    \label{fig:figure_1}
\end{figure}

Despite the fact that human practitioner based diagnosis of respiratory auscultation has been practiced for several hundred years, there exists numerous drawbacks and limitations to this process. Apart from the major limitations of the interpretation of observations being prone to human error and the experience of the practitioner, long-term monitoring of a single patient's recordings or comparisons between different patients is practically infeasible \cite{hsu2021benchmarking}. This motivates the need for an automated lung sound event detection framework where continuous lung sound recordings can be segmented into discrete events.  

In spite of this need, only a handful of studies \cite{hsu2021benchmarking, messner2018crackle, hsiao2020breathing, liu2017respiratory, jacome2019convolutional} have explored the use of deep learning techniques for event detection in lung sounds \cite{hsu2021benchmarking}, and the majority of the literature \cite{gurung2011computerized, sanchez2013computer, pham2021cnn, altan2019deep, shuvo2020lightweight} has simply focused on distinguishing healthy participants from abnormal patients. Furthermore, existing lung event detection frameworks exhibit a lack of generalisation due to data scarcity. Their performance is limited to a small number of diseases/event types, and they have only been evaluated using a single dataset. 

In addressing these research gaps we propose a lightweight, yet robust, simplified and completely interpretable framework for lung sound event detection. We observe that recurrent neural networks such as Long Short-Term Memory (LSTM) \cite{hochreiter1997long} networks, Gated recurrent units (GRUs) \cite{cho2014learning}, and attention based frameworks \cite{fernando2019heart} are a popular choice when modelling temporal relationships within the auscultation recordings. Deviating from this common norm and seeking to develop a lightweight, simplified architecture, we utilise Temporal Convolution Networks (TCNs) \cite{lea2017temporal} as the building block of the proposed framework. 

Through our extensive evaluations we illustrate that a multi-branch TCN architecture can discover temporal relationships at different granularities, which can be missed by attention based recurrent architectures. Moreover, we illustrate how these pieces of identified salient information can be effectively fused such that a lightweight classifier can generate informed decisions without being overwhelmed by a multitude of information. 

The proposed framework is illustrated in Fig. \ref{fig:figure_1}. We extract consecutive overlapping windows from lung recordings and pass these through our feature extraction pipeline. The resultant features are processed by the proposed network, which outputs the event category of the observed input window. Critically, the proposed architecture is simple and completely interpretable, which has been overlooked in existing lung auscultation literature. To this end, we propose a complete framework to interpret the operations of all the components of the proposed architecture. Despite the fact that we are using existing interpretation algorithms in our interpretation framework, such an end-to-end interpretation pipeline is itself novel. We evaluate our system on three challenging datasets (2 public and one in-house) for the identification of numerous auscultation events including, inhalation, exhalation, crackles, wheeze, stridor, and rhonchi, and achieve state-of-the-art results in all evaluations. The main contributions of this work can be summarised as follows:
\begin{itemize}
    \item We propose a novel lung event detection framework which is simple and light-weight in it's design, but is capable of achieving  state-of-the-art results on multiple challenging benchmarks for the detection of event categories including, inhalation, exhalation, crackles, wheeze, stridor, and rhonchi.
    \item We identify the importance of the multi-branch TCN architecture and propose an effective way to combine information from these branches.
    \item We demonstrate how end-to-end interpretations can be generated for the proposed framework, allowing a practitioner to understand what features within the input are leading to the final classification decision. Such interpretations increase the trust in the system, potentially enabling its clinical use in hospital settings.
\end{itemize}

\section{Related Work}

Since its advent, deep learning has revolutionised the filed of machine learning and numerous works have exploited the computational power of deep learning for the detection of abnormality in lung sound recordings. For instance, in \cite{demir2020convolutional} the authors investigate the use of a CNN-SVM hybrid architecture for patient wise classification of abnormalities in lung sound recordings. Similarly in \cite{pham2021cnn} the authors utilise a CNN network coupled with a Mixture of Experts (MoE) as the classifier. Specifically the MoE block contains a series of classifiers and a gating mechanism to automatically decide which classifier should to be applied to which input region. The works \cite{shuvo2020lightweight, minami2019automatic} use 2D CNN architectures for abnormality detection, while the authors of \cite{altan2019deep} use Deep Belief Networks as their classification framework and \cite{messner2020multi} explore the use of a CNN-RNN hybrid model. Despite these extensive investigations, these works have limited practical applications as they only focus on distinguishing healthy participants from abnormal patients. The utility of these methods, and the potential diagnostic outcomes, can be improved by temporally localising the events within the lung sound recordings \cite{hsu2021benchmarking}, which we refer to as ``lung sound event detection''. 

Considering the literature on lung sound event detection, in \cite{jacome2019convolutional} the authors adapt the Faster R-CNN \cite{ren2016faster} framework for lung event detection using a spectrogram representation of the lung recording. Specifically, the region proposal network of the Faster R-CNN provides proposals for inhale and exhale breathing phase locations in the spectrogram and the classifier network assigns the relevant event labels to those proposals. However, considering the large trainable parameter space of the Faster R-CNN the authors could not train the network using lung data from scratch. Hence, they utilised a pre-trained ResNet101 \cite{he2016deep} architecture as the Faster R-CNN backbone network. Furthermore, they have performed post-processing on the network predictions based on author defined assumptions. 

Another ResNet based transfer learning model is proposed in \cite{hsiao2020breathing}. Similar to \cite{jacome2019convolutional} the authors utilise a pre-trained ResNet101 to extract features from the input spectrogram. However, an attention based LSTM pipeline is used to generate event labels. The spectrograms of consecutive windows are sequentially passed through the network, which are encoded using the ResNet. By using the embedding of the current spectrogram and the previous embeddings, the LSTM then generates the event label of the current input. 

The use of Recurrent Neural Network (RNN) is also proposed in \cite{messner2018crackle}. The authors use a bidirectional GRU model which operates directly on the extracted Mel Frequency Cepstral Coefficients (MFCCs) and Spectrogram features. The authors simultaneously predict the breathing phase (i.e inhale/exhale) and the presence/absence of Crackle. However, their evaluations are limited to a private in-house dataset and their model can only detect one category of adventitious sounds. 

A comprehensive evaluation of different off the shelf deep learning architectures is presented in \cite{hsu2021benchmarking}. The authors have benchmarked different RNNs (including LSTM and GRU) and CNN-RNN hybrids (including CNN-LSTM, CNN-GRU, CNN-BiLSTM, and CNN-BiGRU) for the lung sound event detection task. Their evaluations revealed that the CNN-RNN hybrids perform slightly better compared to RNN only methods. But the authors have acknowledged that their performance is inconsistent across numerous event categories. Most importantly, they observe that the performance is heavily influenced by the smaller and imbalanced datasets which are available for the lung sound event detection task. This motivates the need for a lightweight model which can learn from limited data, but have sufficient capacity to model temporal relationships across different time intervals. Furthermore, the authors of \cite{hsu2021benchmarking} have observed that unusual breathing patterns such as shallow/fast breathing and different breathing rates in the recordings, contribute to the model's erroneous predictions. This clearly illustrates the need to capture temporal relationships within the input at different granularities. For instance, if the model is sequentially processing the input time-series (such as in LSTMs and GRUs) temporal relationships that exist across longer timescales can be missed \cite{fernando2018tree, fernando2020neural}. 

Leveraging these observations and motivated by the limitations of the existing literature, we propose a lightweight yet robust deep learning framework for lung sound event detection. We adapt lightweight Temporal Convolution Networks in place of parameter heavy RNN models and CNN-RNN hybrids. Despite being a popular choice in event detection \cite{hsiao2020breathing, fernando2019heart}, we observe that soft-attention mechanisms dramatically increase the parameter space of the network, and hence are not suitable when learning from limited data. To this end we propose an effective feature fusion strategy which propagates only the most salient information to the classifier. 

\section{Methodology}

In this section we first provide a brief introduction to temporal convolutions, illustrating their limitations and motivating the need for a multi-branch Temporal Convolution Network (TCN) architecture. We then illustrate the proposed fusion strategy to fuse the information captured by the distinct branches and the details of the classifier which outputs the event labels. 

\subsection{Temporal Convolution}
Recurrent architectures such as LSTMs and GRUs are outperformed by TCNs for numerous sequence modelling tasks \cite{lea2017temporal, matthewdavies2019temporal} due to their improved memory retention. Unlike LSTMs and their counterparts which maintain a fixed size memory, TCNs can attend to the entire sequence when extracting temporal relationships. Furthermore, the removal of recurrency leads to reduced computational cost. However, the effectiveness of TCNs relies on the amount of temporal connections that the kernel can capture. In this sub-section we illustrate the motivation behind the proposed multi-branch architecture. 

\begin{figure*}[htbp]
    \centering
    \subfloat[][]{\includegraphics[width=.35\linewidth]{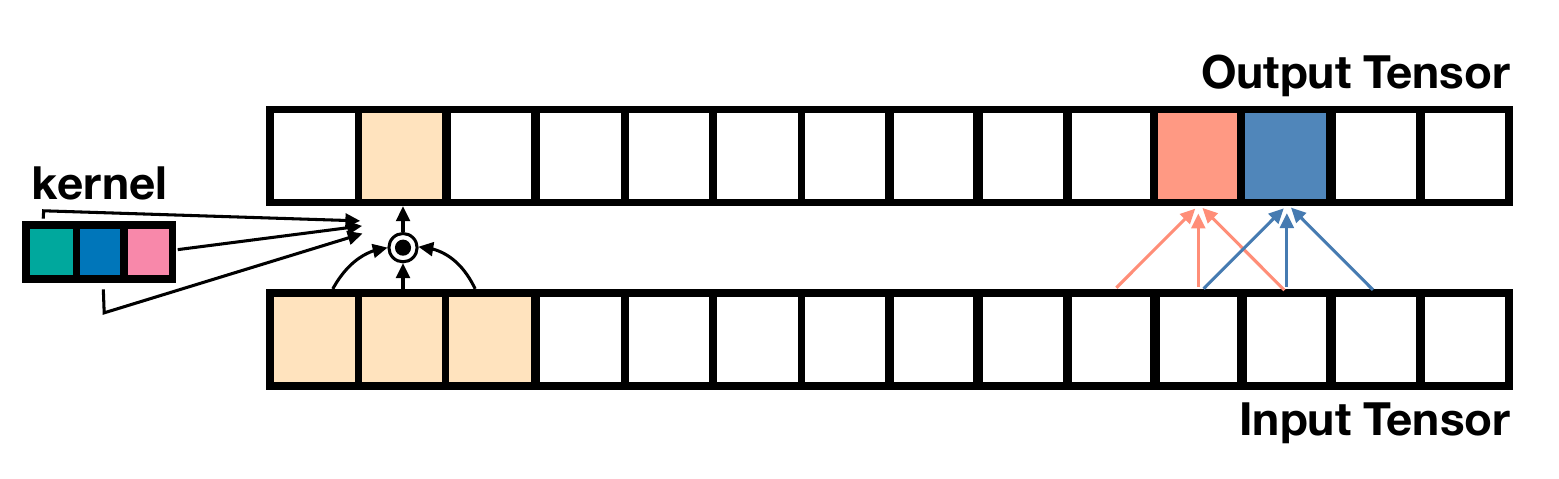}}
    \subfloat[][]{\includegraphics[width=.35\linewidth]{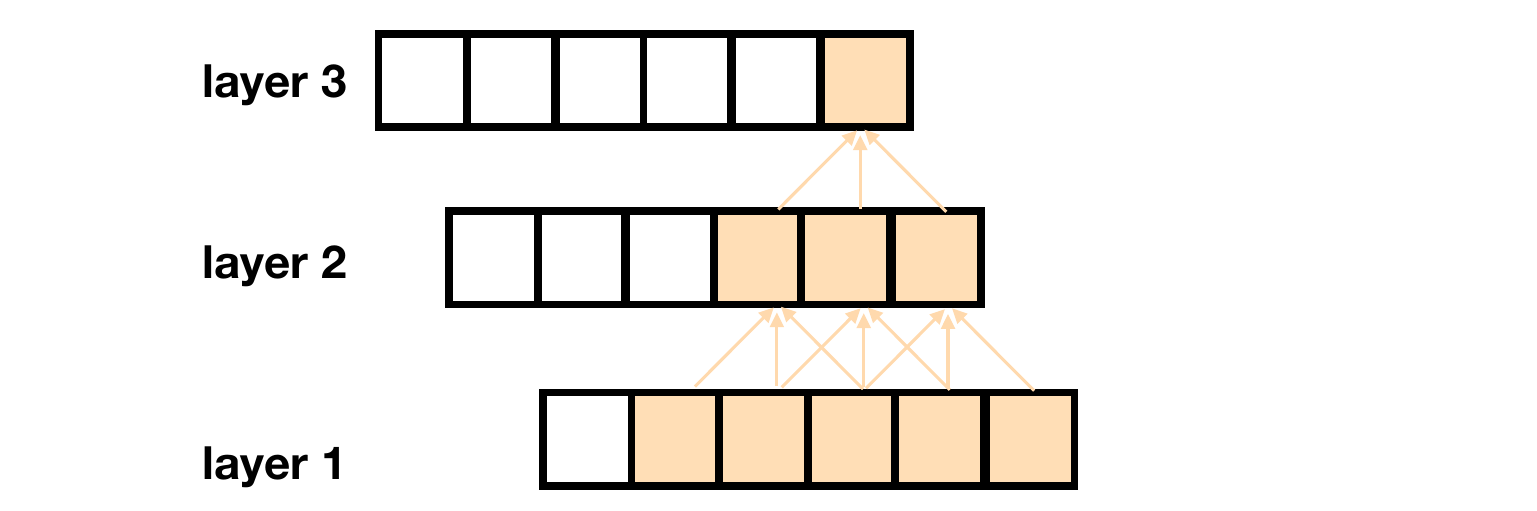}}
    \subfloat[][]{\includegraphics[width=.3\linewidth]{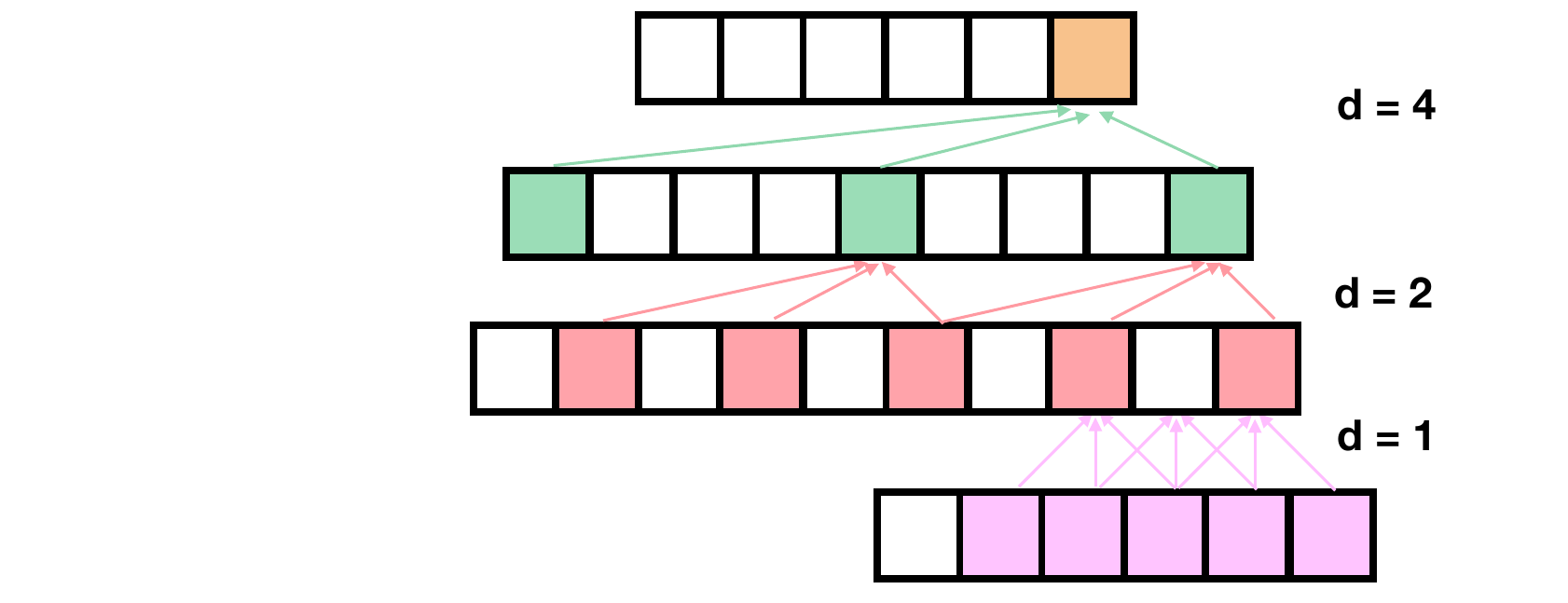}}
    \caption{Illustration of temporal convolution operation and dilation. (a) Input to output information propagation in 1D convolution with a kernel size of 3. (b) The receptive field size increases (to 5) when two 1D convolution layers are stacked. (c) Illustration of dilated convolutions with a kernel size of 3 and dilation factors, $d$, of 1, 2 and 4.}
    \label{fig:TCN_intro}
\end{figure*}

Fig. \ref{fig:TCN_intro} (a) visualises the 1D convolution operation where a 3D tensor (of shape [batch size, seq length, input channels]) is converted to an output tensor of shape [batch size, seq length, output channels] using a 2D kernel. The right-hand side of Fig. \ref{fig:TCN_intro} (a) shows how the convolution output is generated for two consecutive elements in the output tensor, where the kernel is of size 3. The output is the dot product between the input elements within the window and the learned kernel weights. Then the kernel is shifted by one element to the right. Hence, in this case the receptive field size is 3. However, when modelling lengthier sequences we require a larger receptive field such that a particular output element is influenced by temporally distant input elements. 

One way of achieving this is through stacking multiple 1D convolution layers. For instance, as shown in Fig. \ref{fig:TCN_intro} (b), by stacking 2 layers with a kernel size of 3 we can obtain a receptive field size of 5. Another way of achieving this is through dilation. The distance between elements in the input sequence which is used to compute one entry in the output sequence is referred to as the dilation. Fig. \ref{fig:TCN_intro} (c) illustrates how dilation can be used to obtain a larger receptive field without the need to stack a large number of individual convolutions. 

However, as illustrated in \cite{lea2017temporal, farha2019ms}, depending on the level of dilation and the number of convolution layers that are stacked together, there can be `holes' in the receptive field such that elements in the input sequence are missed when generating the output sequence. To this end we propose a multi-branch network architecture which has different dilation bases (i.e 1, 2, 3, \ldots) and we exponentially increase the dilation value when moving up the layers in a particular branch. For example, when the dilation base is 2 the first layer has a dilation factor $d = 2^0 = 1$, the second layer $d = 2^1 = 2$, the third layer $d = 2^2 = 4$, etc. This allows different branches to capture temporal relationships at different granularities. 

\begin{figure}
    \centering
    \includegraphics[width=.5\linewidth]{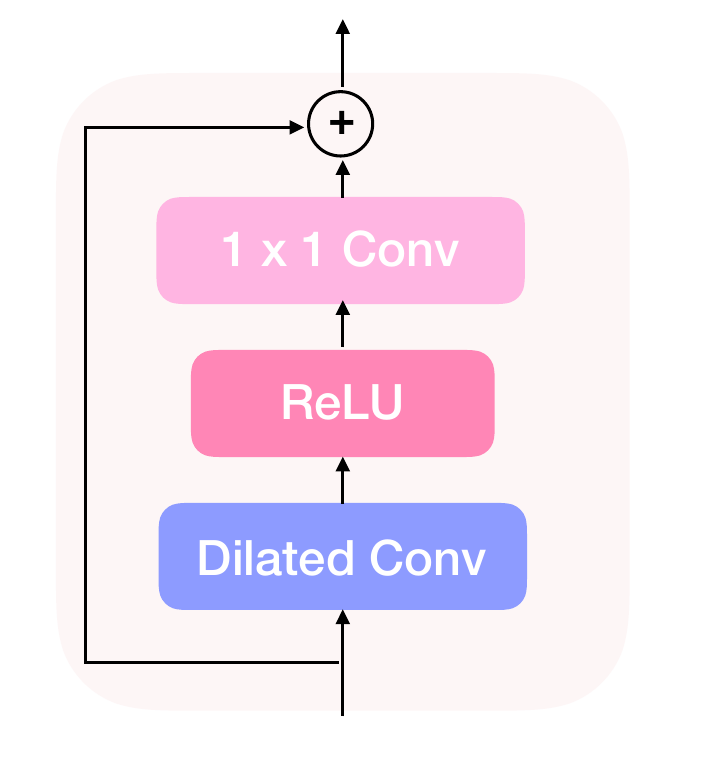}
    \caption{Dilated residual layer utilised in the proposed work.}
    \label{fig:TCN_with_relu}
\end{figure}

Following \cite{farha2019ms} we use dilation convolution layers with residual connections. Fig. \ref{fig:TCN_with_relu} illustrates a residual block. Formally,
\begin{equation}
    \hat{H}_l = \mathrm{ReLU}(W_1 * H_{l-1} + b_1),
\end{equation}
\begin{equation}
    H_l = H_{l-1} + W_2 * \hat{H}_{l} + b_2,
\end{equation}
where $*$ denotes the convolution operation, and $W_1$ and $W_2$ are the weights of the dilated convolution layers (of kernel size 3) and weights of the second convolution layer with kernel size of 1, respectively. $H_l$ is the output of layer $l$, and $b_1$ and $b_2$ are the biases. 

\subsection{Multi-branch TCN}

As outlined earlier, we utilise a multi-branch network to capture different temporal granularities. Formally, let the input recording be denoted by the vector $p$ which contains the audio frames from time instance $0$ to $T_{obs}$, 
\begin{equation}
    p = [X_0, \ldots, X_{T_{obs}}],
\end{equation}
and the feature extractor function, $f$, extracts features from this input such that,
\begin{equation}
    s = f(p).
\end{equation}
Then we pass the feature vector, $s$, through the individual encoders of the $B$ branches which are composed of stacked TCN layers. Let the output $o^b$ denote the output of the encoder of branch $b \in [1, \ldots, b, \ldots, B]$ which consists of $l \in [1, \ldots, l, \ldots, L]$ TCN layers, where the first layer has a dilation factor of $d$, and increases it exponentially from one layer to the next, 
\begin{equation}
    o^b = \mathrm{Enc}^{b, l, d}(s).
\end{equation}

\subsection{Classifier}

The output, $o^b$, is a 3 dimensional tensor of shape [batch size, seq length, output dim]. In order to combine the outputs of the $B$ branches we concatenate them on the 2nd dimension such that the combined tensor, $\hat{o}$, is of shape [batch size, seq length $\times$ $B$, output dim].

\begin{figure}[htbp]
    \centering
    \includegraphics[width=\linewidth]{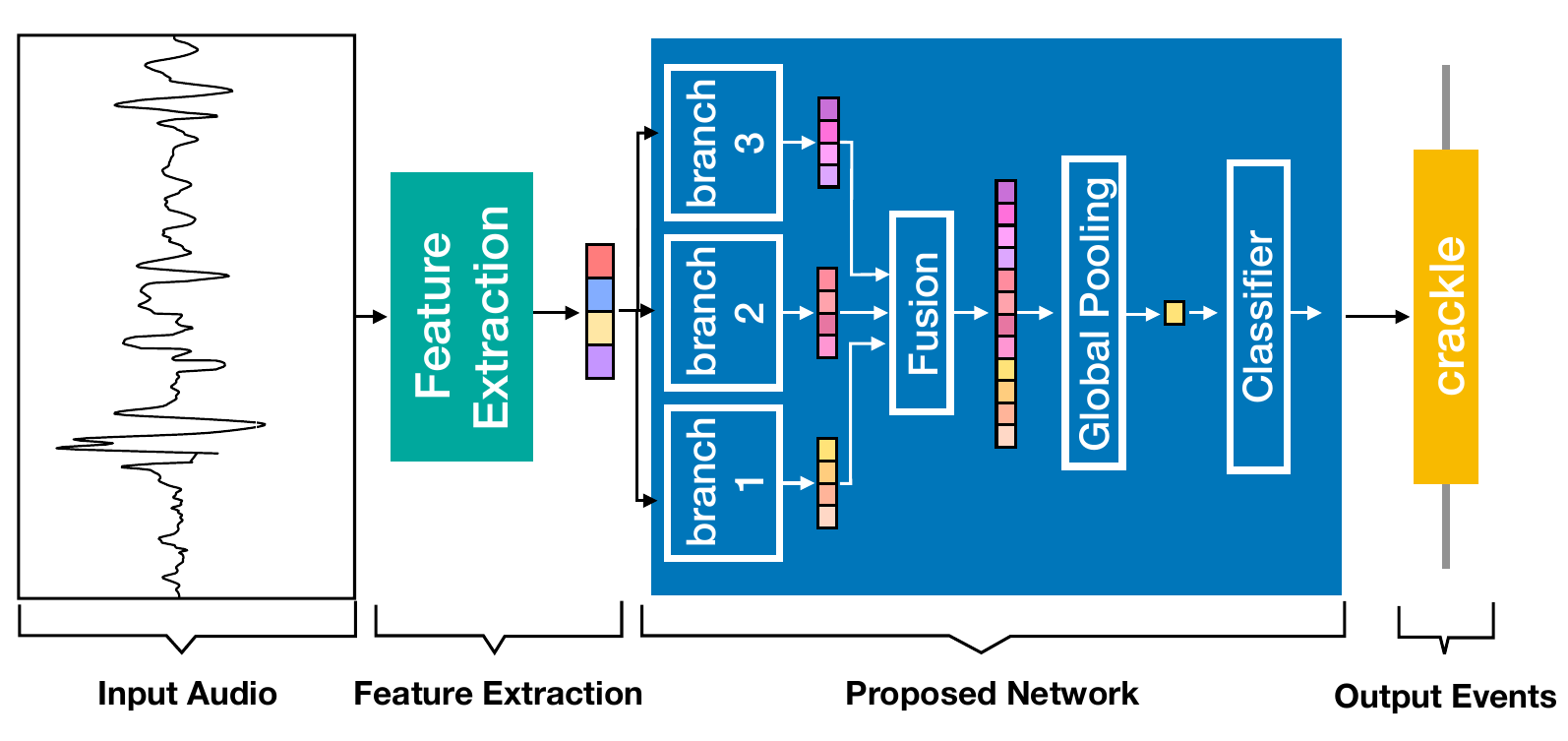}
    \caption{Proposed model architecture: Once the features are extracted from the input lung sound recording they are passed through separate branches, which extract features at different temporal granularities. Resultant feature vectors are fused together using concatenation and global average pooling is applied to obtain a fixed-size 2D vector, which is subsequently passed through a classifier to classify the event category of the input window.}
    \label{fig:model}
\end{figure}

To enable the system to process input audio recordings/segments of variable lengths\footnote{Note that while the system can process sequences of arbitrary lengths, in our evaluations we use 1 second windows} we use a global average pooling operation to extract a fixed size 2D tensor from the concatenated output, $\hat{o}$. We note that global average pooling is performed separately for each feature dimension. Therefore, if we apply feature level concatenation (naive-concatenation), which is performed by concatenating on the third dimension, this would increase the input feature dimension of the classifier. However, our multi-branch architecture generates different feature representations of the same input segment. Therefore, by concatenating features on the second dimension and pooling the features across this entire dimension, we force the branches to propagate only the most salient features to the classifier. Hence only the most important features across the $B$ branches will be activated and the rest of the elements will be inactive. Therefore, the global average pooling operation will focus on these features, significantly reducing the overhead on the classifier. This allows us to use a simple, light-weight 3 layer multi-layer perceptron as our classifier. The overall architecture of the proposed method is visually illustrated in Fig. \ref{fig:model}.

\section{Experiments}
\subsection{Datasets}

\subsubsection{HF Lung V1 Database \cite{hsu2021benchmarking}}
This dataset is captured from 279 patients and comprises 9,765 recordings, each of 15 sec duration. This is the largest publicly available lung sound recording dataset to date. The authors provide inhalation and (34,095) exhalation (18,349) labels, and label adventitious breathing sounds including 13,883 Continuous Adventitious Sounds (CAS) (8,457 wheezes, 686 stridor and 4,740 rhonchi events), and 15,606 Discontinuous Adventitious Sounds (DAS) (all crackles). Following the dataset authors we conducted separate experiments for inhalation, exhalation, CAS and DAS event detection. For training and testing we utilised the splits provided by the dataset authors, comprising 7,809 recordings for training and 1,956 for testing. As the validation set we used 10\% of the training data. 

\subsubsection{ICBHI Respiratory Sound Database \cite{rocha2017alpha}}
This dataset contains 920 annotated recordings with lengths varying from 10 to 90 sec, taken from 126 patients. The dataset has been annotated for respiratory cycles (6,898 cycles in total) and has 1,864 crackles, 886 wheezes and 506 events with both crackles and wheezes. It should be noted that this dataset replicates real life conditions with noisy recordings captured from different equipment (microphone, 3M Littmann stethoscope and Meditron stethoscope), different chest locations (7 locations in total), and spans different age groups (ranging from children, to adults and the elderly). As this dataset does not contain inhalation/exhalation annotations we utilise it only for the detection of adventitious breathing sounds. Specifically, we conducted experiments to detect crackles and wheezes. Following prior works \cite{pham2020cnn, minami2019automatic} we randomly split the data into training and testing where 60\% of the data is used for training and 40\% for testing. We ensure that when sampling the recordings, samples from the same recording fall either to training or testing splits, not both. As the validation data we used 10\% of the training set.

\subsubsection{M3DICINE Lung Sound Dataset}

The M3DICINE Lung Sound Dataset consists of 224 recordings collected from 78 patients using M3DICINE’s Stethee\textregistered{}\footnote{https://www.stethee.com/} Pro 1 device. The recordings were collected from adults and children of different ethnic backgrounds and age groups. The recordings are 20 seconds in length, and are collected from  the intersection of the mid-clavicular line and second intercostal space on the Right anterior surface, as well as posteriorly on the participant at a level lateral to T3 . These locations were selected as they represent the locations of the Left Upper Lobe and Right Upper/Middle Lobes of the lungs, locations that are used routinely as part of respiratory clinical examinations and also in computer aided analysis of lung sounds. We randomly split the data into training and testing where the 70\% of the data is used for training and 30\% for testing. 10\% of the training data is used for validation. We ensure that when sampling the recordings, samples from the same recording fall either to training or testing splits, not both.

\subsection{Pre-processing, Feature Extraction and Data Augmentation}
Following \cite{hsu2021benchmarking}, in all datasets we resampled the recordings to 4kHz and applied a high-pass filter with cut-off frequency 80Hz and filter order of 10 to remove electrical interference and other background noise. We applied a Hamming window of size 4000 (1 sec) and a window shift of 2000 (.5 sec) to extract samples. As the label of the extracted window, we used the majority label of the window. 

As features we extracted Mel Frequency Cepstral Coefficients (MFCCs) by obtaining $13$ static, $13$ delta and $13$ delta-delta coefficients in the fequency range $0-2000$Hz. In addition, we extracted the log mel filter bank energy feature of the window using $26$ filters. All the features are concatenated together, yielding a feature vector of dimension $65$. For all feature extraction processes we used a window length of $0.025$ sec and a step size of $0.01$ sec. Subsequently, min-max normalisation is applied to normalise the features to the range $0$ and $1$. 

As a data augmentation procedure, following \cite{park2019specaugment}, we applied a frequency and time masking technique to randomly mask time slices and frequency bands with a probability of $0.5$. Specifically, we select time and frequency bands, each with $0.5$ probability and randomly select an index to start masking. For frequency masking a window width between 2 and 8 is randomly selected, while for time masking we used a random number of times slice between $5$ and $10$. 

We would like to note that unlike \cite{hsu2021benchmarking}, no post processing steps were applied in our framework. We threshold model predictions by setting values $>$ $0.5$ to $1$ and rest of the values to $0$.

\subsection{Evaluation Metrics}

We use the same evaluation protocol as the authors of \cite{hsu2021benchmarking}. We compare the predictions for each segment with their respective ground truth and evaluate the Jaccard similarity between the ground truth and predicted events. If the Jaccard similarity is greater than $0.5$ we consider it to be a True Positive (TP), if similarity is between $0$ and $0.5$ we consider it a False Negative (FN) and if it is equal to zero we consider it a False Positive (FP). It should be noted that a True Negative (TN) event cannot be defined for this event detection task. We evaluated the Positive Predicted value (PPv), Sensitivity (Se) and F1 score using the following definitions,

\begin{equation}
PPv = \frac{TP}{TP + FP},
\end{equation}

\begin{equation}
Se = \frac{TP}{TP + FN},
\end{equation}

\begin{equation}
F_1 = 2 \frac{PPv . Se}{ PPv + Se}.
\end{equation}

\subsection{Implementation Details}

Implementation of the proposed model is completed using Pytorch (version 1.5.0) and CUDA (10.1), and the model is trained on an NVIDIA GForce RTX 2080 GPU. 

The model contains number of branches, $B$, the number of layers, $l$, in each branch and the number of filters, $k$, of each convolution layer as hyper-parameters. These were evaluated experimentally and were set to $b=3$, $l=3$ and $k=80$. Please refer to Sec. \ref{sec:discussion} for the evaluation details. The kernel size of all the convolution layers is set to 3 and dilation bases are set to 2, 3 and 4 for the 3 branches. The classifier network consist of three multi-layer perceptions with hidden dimensions, 80, 32 and 1, respectively. The proposed model was trained for 200 epochs using binary cross entropy as the objective and the Adam \cite{kingma2014adam} optimiser, using mini-batch of size 64, and a learning rate of $10^{-5}$. 

\subsection{Results}

An evaluation of the proposed framework together with the current state-of-the-art results for the HF Lung V1, ICBHI, and M3dicine Lung datasets is provided in Tab. \ref{tab:main_results}. To the best of our knowledge, apart from \cite{hsu2021benchmarking} none of the existing works have evaluated their lung event detection methods on publicly available data. Therefore, we utilise the LSTM, BiGRU, CNN-GRU and CNN-Bi-GRU models proposed in \cite{hsu2021benchmarking} as our baselines for evaluations. As the implementations of \cite{hsu2021benchmarking} are not publicly available we carefully re-implemented these models to provide comparison results for the ICBHI and M3dicine Lung datasets (results on HF Lung V1 are taken from \cite{hsu2021benchmarking}). To illustrate the robustness of the models, we randomly initialised the network weights and ran each experiment 3 times, and report the average metric values along with the standard deviation.

\begin{table}[htbp]
\caption{Evaluation Results of the HF Lung V1, ICBHI, and M3DICINE Lung datasets for the detection of Inhalation, Exhalation, Continuous Adventitious Sounds (CAS), Discontinuous Adventitious Sounds (DAS), Crackles and Wheezes events. We report Positive Predicted value (PPv), Sensitivity (Se) and F1 scores.}
\label{tab:main_results}
\centering
\resizebox{\linewidth}{!}{
\begin{tabular}{|c|c|c|c|c|c|}
\hline
\multirow{2}{*}{Dataset}       & \multirow{2}{*}{Task}     & \multirow{2}{*}{Models} & \multicolumn{3}{c|}{Results} \\ \cline{4-6} 
                               &                           &                         & PPv      & Se      & F1      \\ \hline
\multirow{20}{*}{HF Lung V1}   & \multirow{5}{*}{Inhalation}        & LSTM \cite{hsu2021benchmarking}                   &   0.890       & 0.664        &  0.761       \\ \cline{3-6} 
                               &                           & Bi-GRU \cite{hsu2021benchmarking}                 &  0.898        & 0.800        &  0.862       \\ \cline{3-6} 
                               &                           & CNN-GRU \cite{hsu2021benchmarking}                & 0.906         &  0.742       &  0.820       \\ \cline{3-6} 
                               &                           & CNN-Bi GRU \cite{hsu2021benchmarking}             &  0.898        & 0.812        & 0.862        \\ \cline{3-6} 
                               &                           & Proposed                & \textbf{0.998 \textpm 0.021}         & \textbf{0.892 \textpm 0.015}        & \textbf{0.942 \textpm 0.017}        \\ \cline{2-6} 
                               & \multirow{5}{*}{Exhalation}        & LSTM \cite{hsu2021benchmarking}                   & 0.561         & 0.456        & 0.570        \\ \cline{3-6} 
                               &                           & Bi-GRU \cite{hsu2021benchmarking}                 & 0.713         & 0.617        & 0.709        \\ \cline{3-6} 
                               &                           & CNN-GRU \cite{hsu2021benchmarking}                & 0.629         & 0.516        & 0.620        \\ \cline{3-6} 
                               &                           & CNN-Bi GRU \cite{hsu2021benchmarking}             & 0.693         & 0.600        & 0.685        \\ \cline{3-6} 
                               &                           & Proposed                & \textbf{0.976 \textpm 0.011}         & \textbf{0.729 \textpm 0.008}        & \textbf{0.835 \textpm 0.032}        \\ \cline{2-6} 
                               & \multirow{5}{*}{CAS}      & LSTM   \cite{hsu2021benchmarking}                 &  0.120        & 0.095        & 0.122        \\ \cline{3-6} 
                               &                           & Bi-GRU \cite{hsu2021benchmarking}                 &  0.237        & 0.227        & 0.256        \\ \cline{3-6} 
                               &                           & CNN-GRU \cite{hsu2021benchmarking}                &  0.556        & 0.402        & 0.498       \\ \cline{3-6} 
                               &                           & CNN-Bi GRU \cite{hsu2021benchmarking}              &  0.508        & 0.463        & 0.516        \\ \cline{3-6} 
                               &                           & Proposed                &  \textbf{0.895 \textpm 0.024}        & \textbf{0.627 \textpm 0.053}        & \textbf{0.737 \textpm 0.012}        \\ \cline{2-6} 
                               & \multirow{5}{*}{DAS}      & LSTM  \cite{hsu2021benchmarking}                  &  0.699        & 0.485        & 0.591        \\ \cline{3-6} 
                               &                           & Bi-GRU  \cite{hsu2021benchmarking}                &  0.765        & 0.638        & 0.714        \\ \cline{3-6} 
                               &                           & CNN-GRU \cite{hsu2021benchmarking}                &  0.709        & 0.539        & 0.646        \\ \cline{3-6} 
                               &                           & CNN-Bi GRU \cite{hsu2021benchmarking}             &  0.718        & 0.626        & 0.700        \\ \cline{3-6} 
                               &                           & Proposed                &  \textbf{0.997 \textpm 0.032}        & \textbf{0.787 \textpm 0.025}        & \textbf{0.880 \textpm 0.034}        \\ \hline \hline
\multirow{10}{*}{ICBHI}        & \multirow{5}{*}{Crackles} & LSTM \cite{hsu2021benchmarking}                   &  0.646        & 0.507        & 0.568        \\ \cline{3-6} 
                               &                           & Bi-GRU \cite{hsu2021benchmarking}                 & 0.667         & 0.556        & 0.606        \\ \cline{3-6} 
                               &                           & CNN-GRU \cite{hsu2021benchmarking}                & 0.646         & 0.507        & 0.568       \\ \cline{3-6} 
                               &                           & CNN-Bi GRU \cite{hsu2021benchmarking}             &  0.805        & 0.648        &  0.718        \\ \cline{3-6} 
                               &                           & Proposed                & \textbf{0.938 \textpm 0.023}         & \textbf{0.874 \textpm 0.031}        & \textbf{0.905 \textpm 0.026}        \\ \cline{2-6} 
                               & \multirow{5}{*}{Wheezes}  & LSTM  \cite{hsu2021benchmarking}                  & 0.760         & 0.497        & 0.601        \\ \cline{3-6} 
                               &                           & Bi-GRU \cite{hsu2021benchmarking}                 & 0.804         &  0.556       & 0.657        \\ \cline{3-6} 
                               &                           & CNN-GRU  \cite{hsu2021benchmarking}               & 0.815         & 0.537        & 0.647        \\ \cline{3-6} 
                               &                           & CNN-Bi GRU  \cite{hsu2021benchmarking}            & 0.792         & 0.518       & 0.627        \\ \cline{3-6} 
                               &                           & Proposed                & \textbf{0.994 \textpm 0.035}        & \textbf{0.858 \textpm 0.020}        & \textbf{0.921 \textpm 0.019}        \\ \hline \hline
\multirow{5}{*}{M3DICINE Lung} & \multirow{5}{*}{Exhalation}        & LSTM  \cite{hsu2021benchmarking}                  &    0.623      &   0.702      & 0.660        \\ \cline{3-6} 
                               &                           & Bi-GRU \cite{hsu2021benchmarking}                 &   0.667       &   0.735      &  0.699       \\ \cline{3-6} 
                               &                           & CNN-GRU  \cite{hsu2021benchmarking}               &  0.769        &  0.727       &  0.748       \\ \cline{3-6} 
                               &                           & CNN-Bi GRU  \cite{hsu2021benchmarking}            &   0.710       &  0.831       &  0.766       \\ \cline{3-6} 
                               &                           & Proposed                &    \textbf{0.984 \textpm 0.031 }    &  \textbf{0.940 \textpm 0.023 }   &  \textbf{ 0.962 \textpm  0.020 }  \\ \hline
\end{tabular}}
\end{table}

Analysing the results, it is clear that the proposed model has been able to outperform the recurrent and CNN models by a significant margin. We would like to highlight the number of trainable parameters in the baseline models LSTM (300,609), Bi-GRU (552,769), CNN-GRU (2,605,249) and CNN-Bi GRU (5,240,513), whereas the proposed method only contains 276,225 trainable parameters. We would like to emphasise the fact that our lightweight model has been able to achieve significantly higher results compared to these baselines. Furthermore, its performance is consistent across all considered event categories, while we observe significant fluctuations in the performance of the baselines. We believe this behaviour of the baselines results from the high number of trainable parameters, which leads to over-fitting (and poor generalisation) when trained using smaller datasets. In the following section we quantitatively and qualitatively analyse the contributions from each of the proposed innovations, which allow our model to achieve this level of performance. 

\section{Discussion}
\label{sec:discussion}
In this section we evaluate the contributions from the proposed multi-branch TCN architecture and the importance of the proposed feature fusion strategy. Moreover, we illustrate the proposed model interpretation algorithmic framework along with the visualisations for randomly selected samples. 

\subsection{Importance of Multi-branch Architecture}

We evaluate the performance of three ablation models, with the number of branches set to 1, 2 and 4. Dilation bases for the networks are set to 2, (2, 3) and (2, 3, 4 and 5) respectively. Their evaluation results along with the performance of the proposed method (using 3 branches) is presented in Tab. \ref{tab:ablation_1}. For the single branch model we simply take the global average across the time dimension. Note that evaluation results here are presented for the validation set of the HF Lung V1 dataset for the inhale detection task. 

\begin{table}[htbp]
\caption{The effect of number of branches $B$ on the overall accuracy. The results are presented for the validation set of HF Lung V1 dataset for the inhale detection task.}
\label{tab:ablation_1}
\centering
\resizebox{\linewidth}{!}{%
\begin{tabular}{|c|c|c|c|}
\hline
\multirow{2}{*}{Models} & \multicolumn{3}{c|}{Results} \\ \cline{2-4} 
                        & PPv      & Se      & F1      \\ \hline
1 Branch                    & 0.937 \textpm 0.027        & 0.880 \textpm 0.045       &  0.908 \textpm 0.051      \\ \hline
2 Branch                  & 0.991 \textpm 0.036         &  0.877 \textpm 0.029      &   0.931 \textpm 0.038      \\ \hline
3 Branch (Proposed)                & \textbf{0.998 \textpm 0.034}         & \textbf{0.905 \textpm 0.013}        & \textbf{0.966 \textpm 0.025}        \\ \hline
4 Branch              &  0.995 \textpm 0.033        &  0.889 \textpm 0.042      &  0.939  \textpm 0.038     \\ \hline
\end{tabular}}
\end{table}

When analysing the accuracy gain between the single branch and multi-branch architectures, we observe the importance of considering dependencies across multiple temporal granularities. In particular, we observe that the performance gain is maximised in the 3 branch setting and slightly decreases when it is increased above 3. We believe this behaviour is due to the rapid increase in size of the concatenated feature vector, which could confuse the classifier when the information is redundant. 

\begin{figure*}[htbp]
    \centering
    \subfloat[][]{\includegraphics[width=.495\linewidth]{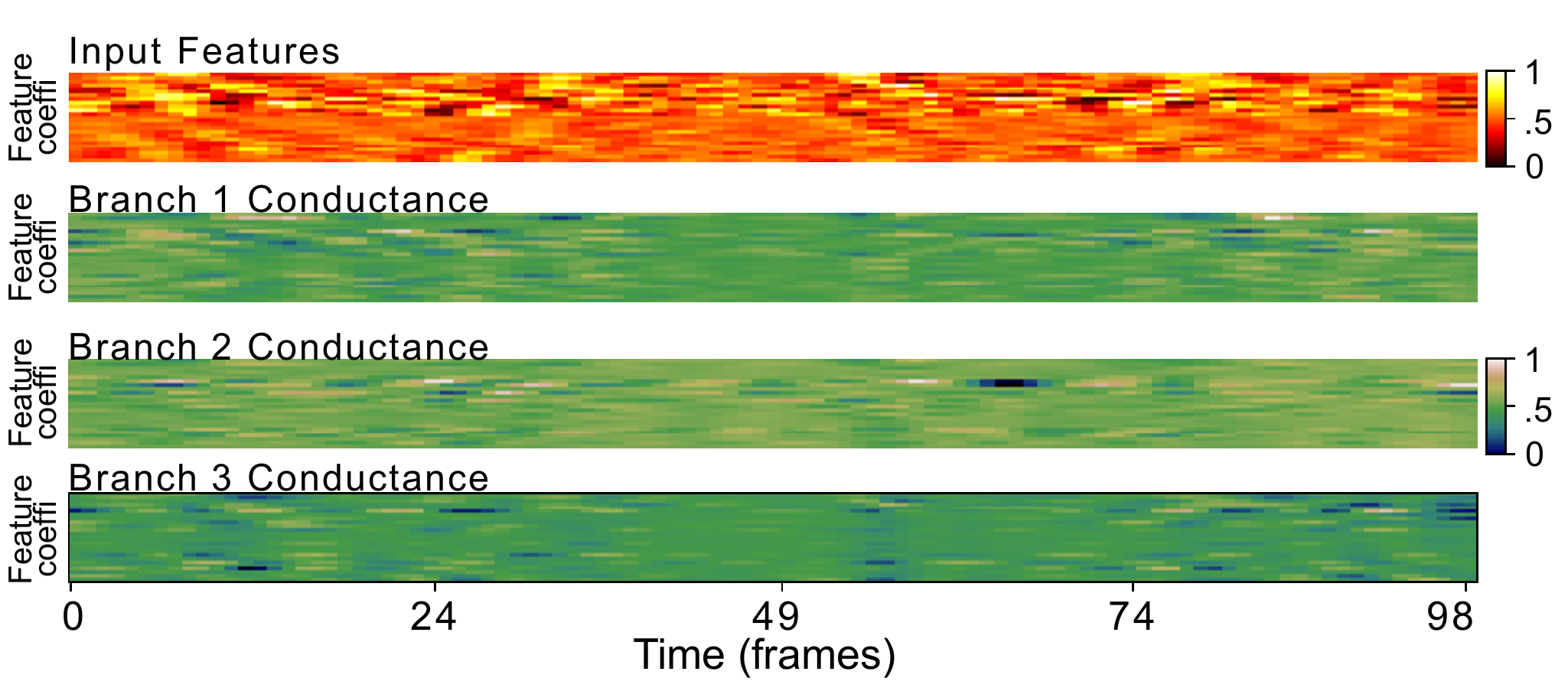}}
    \subfloat[][]{\includegraphics[width=.495\linewidth]{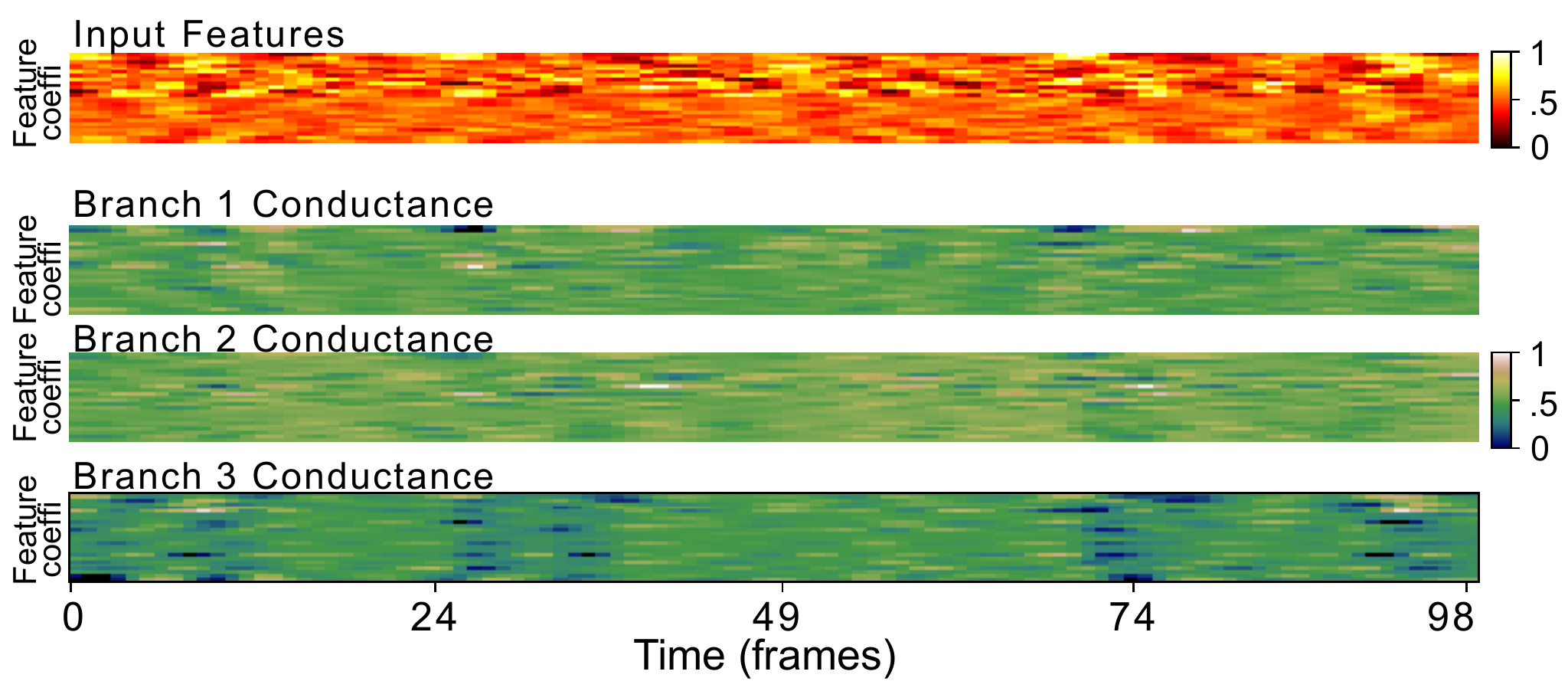}}
    \caption{Neuron Conductance \cite{dhamdhere2018important} for two randomly chosen windows from the `steth\_20190801\_10\_54\_04.wav' recording from the test set of the HF Lung V1 dataset. We visualise the input features along with the conductance of the last layers of the 3 branches.}
    \label{fig:neuron_conductance}
\end{figure*}

Moreover, we visualise the neuron conductance \cite{dhamdhere2018important} of the 3 branches for 2 randomly chosen examples from the HF Lung V1 dataset. The conductance shows the flow of information from the input to a particular hidden unit. By aggregating it across all the hidden units of a particular layer we can identify which regions of the input are activated (i.e conductance with respect to each input feature), and therefore the attributed output is of the same size as the input. Fig. \ref{fig:neuron_conductance} illustrates the neuron conductance of the last layers of the 3 branches of the proposed model for two examples, and demonstrates that the salient information identified by the 3 branches varies. Recall that each branch has different dilation bases, which allows them to see different temporal dependencies. Therefore, we see different input regions being activated by different neurons among these branches. 

\begin{figure}[htbp]
    \centering
    \subfloat[][]{\includegraphics[width=.495\linewidth]{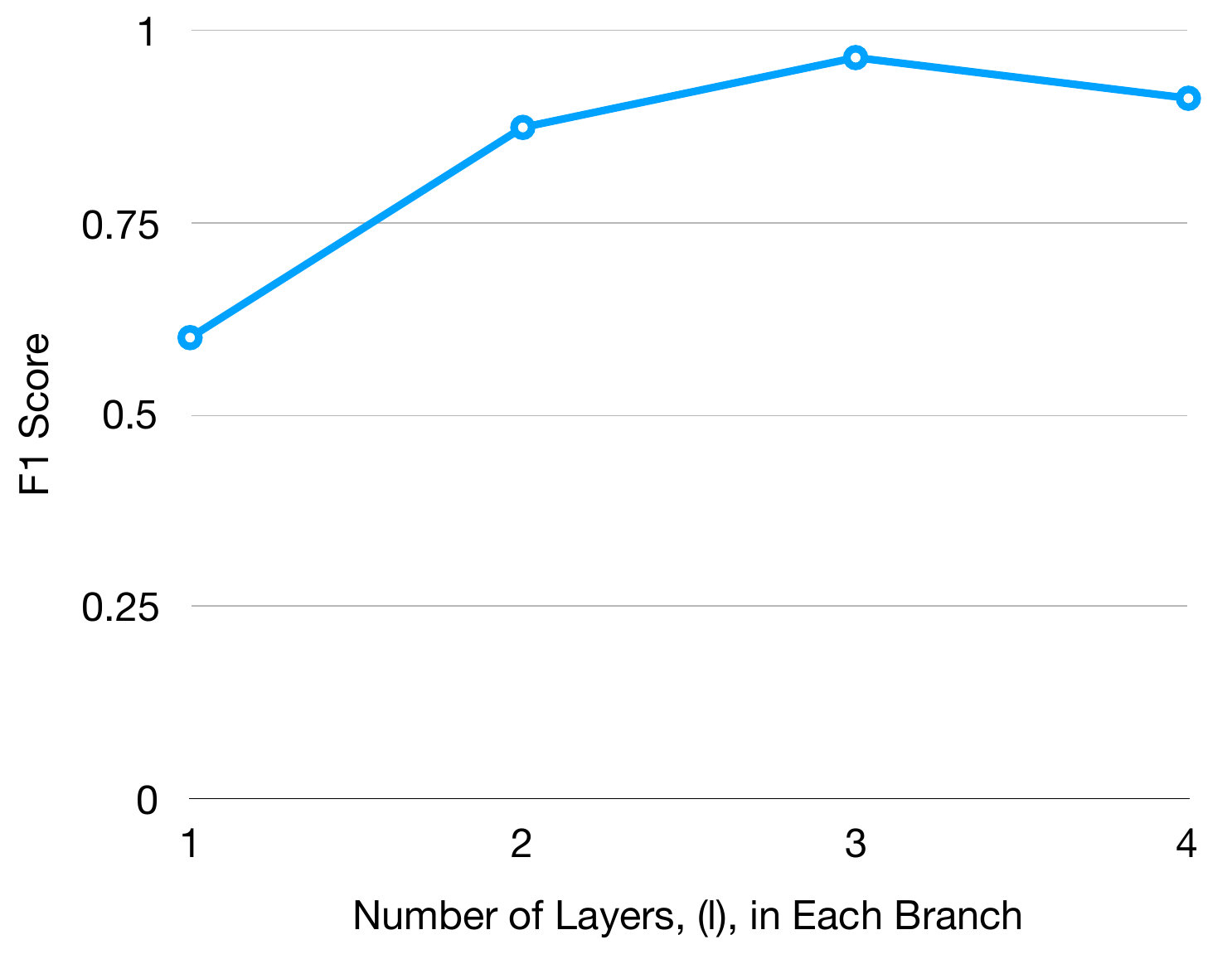}}
    \subfloat[][]{\includegraphics[width=.495\linewidth]{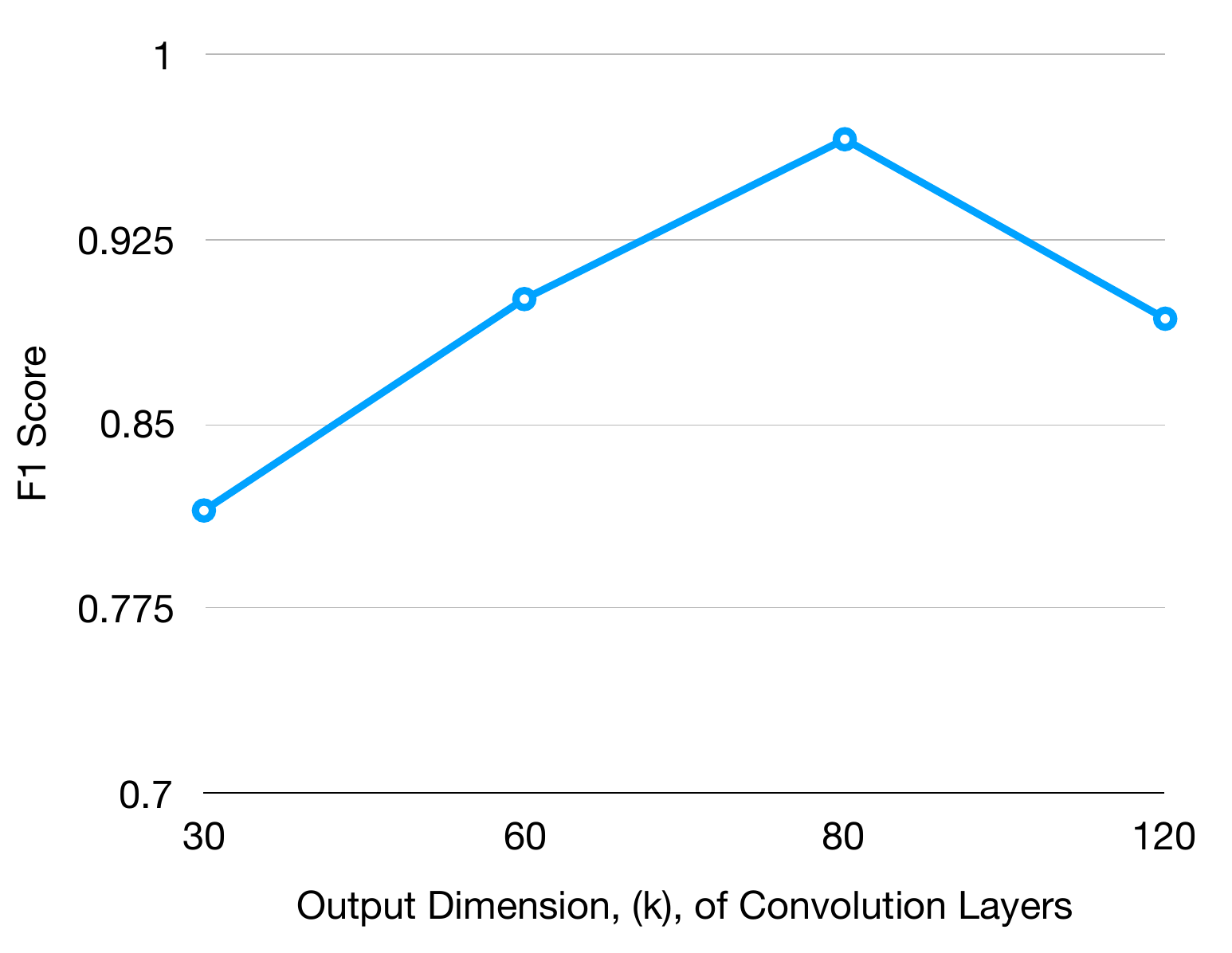}}
    \caption{Hyper-Parameter Evaluations: (a) Change in F1 score against number of layers, $l$, in each branch. (b) Change in F1 score against output dimension, $k$, of convolution layers. Results are for the validation set of of HF Lung V1 dataset for the inhale detection task, using three branches in the network.}
    \label{fig:hyper_parameters}
\end{figure}

Similarly we evaluated the optimal number of layers, $l$, in each branch and the output dimension, $k$, of convolution layers, which are presented in Fig. \ref{fig:hyper_parameters} (a) and (b), respectively. Note that for this evaluation we fixed $B = 3$ and used the validation set of of HF Lung V1 dataset for the inhale detection task. The plots show the F1 score. 

\subsection{Importance of the Proposed Feature Fusion Mechanism}

Next, we investigate the importance of the proposed feature fusion mechanism where we concatenate the features of the $3$ branches on the time axis (i.e 2nd dimension), and apply global average pooling across this dimension to extract a fixed 2D tensor. 

Firstly, we evaluate the performance of naive concatenation (i.e concatenating across the feature dimension, the 3rd dimension) and then taking global average pooling across the 2nd dimension. We evaluate two other attention based feature fusion mechanisms. In the first model we apply multi-head self-attention \cite{vaswani2017attention} to extract features from the naive concatenated tensor where the number of attention heads is set to 4. In the second model we use two levels of self-attention, where individual attention weights were applied to individual branches to extract salient features from those branches. Then these `enhanced features' are concatenated together on the feature dimension (i.e 3rd dimension). Then we apply another level of attention to summarise this concatenated feature across the time axis (i.e convert it to a 2D tensor). The evaluations of these ablation models along with the proposed method for the HF Lung V1 dataset's inhale detection task are presented in Tab. \ref{tab:ablation_2}

\begin{table}[htbp]
\caption{Comparison between the proposed feature fusion mechanism, naive concatenation and attention based feature fusion. The results are presented for the validation set of HF Lung V1 dataset for the inhale detection task.}
\label{tab:ablation_2}
\centering
\resizebox{\linewidth}{!}{%
\begin{tabular}{|c|c|c|c|}
\hline
\multirow{2}{*}{Models} & \multicolumn{3}{c|}{Results} \\ \cline{2-4} 
                        & PPv      & Se      & F1      \\ \hline
naive concat                    & 0.927 \textpm 0.036        & 0.861  \textpm 0.018      & 0.893  \textpm 0.026      \\ \hline
Attention 1                  &  0.936 \textpm 0.042       &  0.878 \textpm 0.027 \textpm 0.037     & 0.906 \textpm 0.031       \\ \hline
Attention 2                &  0.938 \textpm 0.021       & 0.877 \textpm 0.056       & 0.906 \textpm 0.047       \\ \hline
Proposed                & \textbf{0.998 \textpm 0.034}         & \textbf{0.905 \textpm 0.013}        & \textbf{0.966 \textpm 0.025}         \\ \hline
\end{tabular}}
\end{table}

\begin{figure*}[htbp]
    \centering
    \subfloat[][]{\includegraphics[width=.53\linewidth]{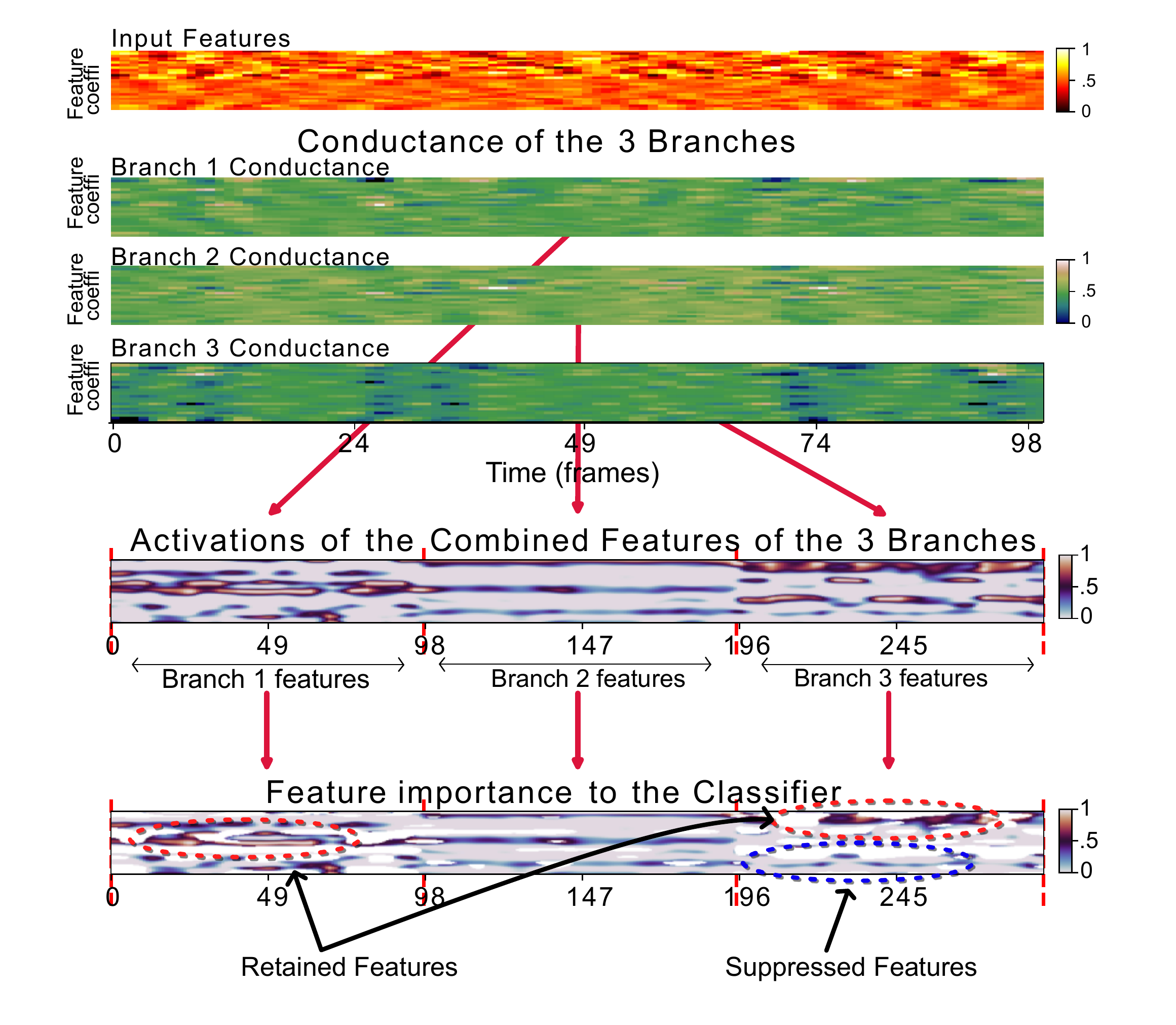}}
    \subfloat[][]{\includegraphics[width=.42\linewidth]{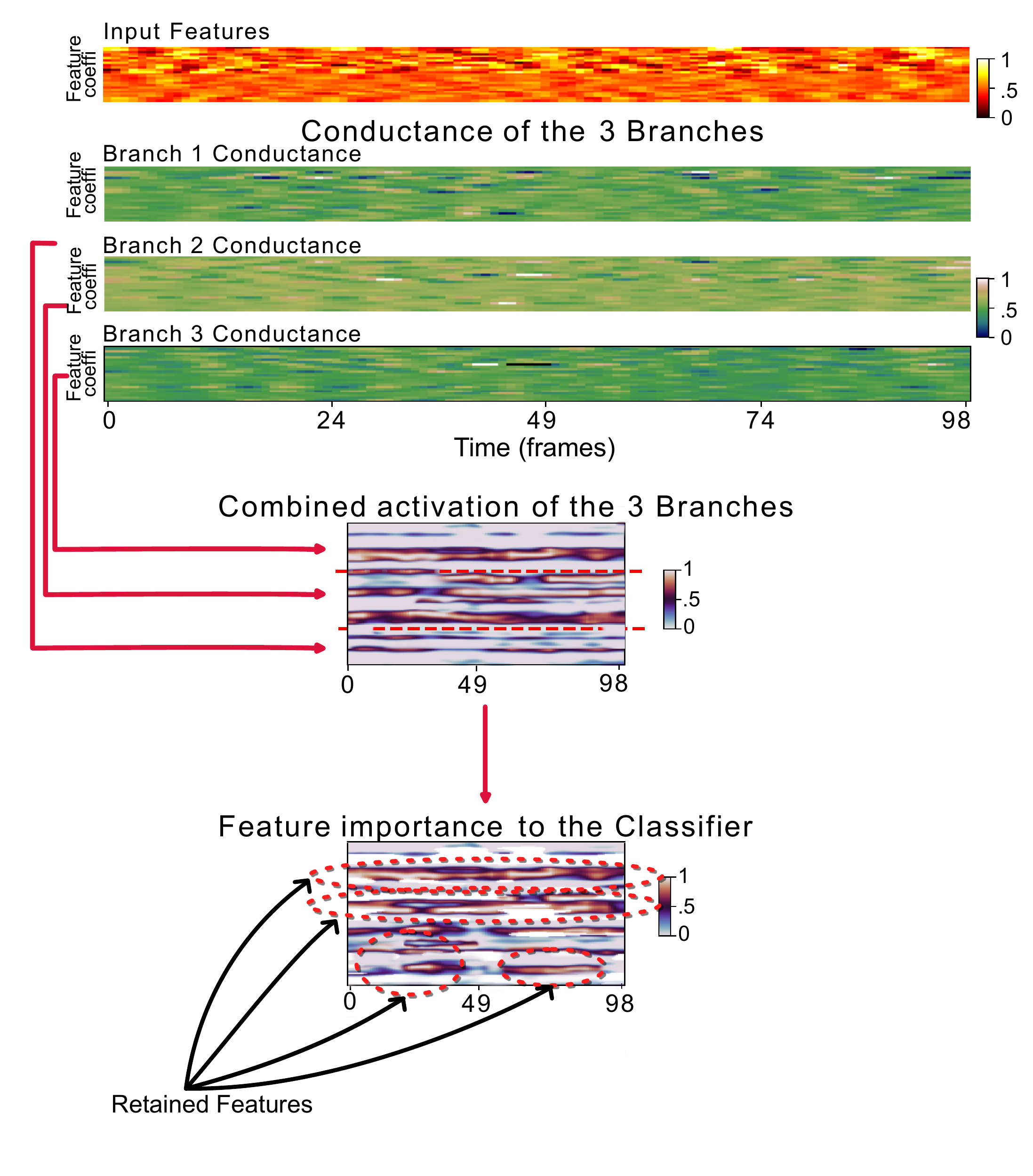}}
    \caption{The activation of the combined feature and its saliency to the classifier (generated using the DeepLift \cite{shrikumar2017learning}) for a randomly chosen window of the recording  `steth\_20190801\_10\_54\_04.wav' from the test set of the HF Lung V1 dataset. In (a) we visualise the saliency for the proposed feature concatenation mechanism and in (b) for the attention based concatenation (attention method 2).}
    \label{fig:fusion}
\end{figure*}

We observe poor performance from both attention based ablation models as well as the naive concatenation mechanism. We believe this is due to the large amount of features present in the 3 branches, which confuses the classifier as the attention/naive concatenation mechanism fails to effectively filter out unnecessary information. Despite the high capacity of the attention based models we do not have sufficiently large datasets to train such high capacity models, yielding poor performance on unseen data. However, in the proposed fusion method we force the branches to pass only the most salient information as we pool across time $\times$ number of branches when passing the information to the classifier. Therefore, the classifier has less information to process, and less chance to get confused by redundant information. To better illustrate this we obtained activations of the concatenated output of the 3 branches along with the attribution of this feature (feature importance to the classifier) generated using the DeepLift \cite{shrikumar2017learning} algorithm. In the visualisations presented in Fig. \ref{fig:fusion} (a) we observe only a small portion of the concatenated tensor is activated and the pooling operation has suppressed most of the features, resulting in a simplified task for the classifier. In Fig. \ref{fig:fusion} (b) we present the same visualisation for the attention model 2. Despite the self-attention based feature selection mechanism we observe that most of the features are activated, hence passing a myriad of information (much of which is redundant) to the classifier. 

\subsection{End-to-End Interpretation Pipeline}

\begin{figure}
    \centering
    \includegraphics[width=.95\linewidth]{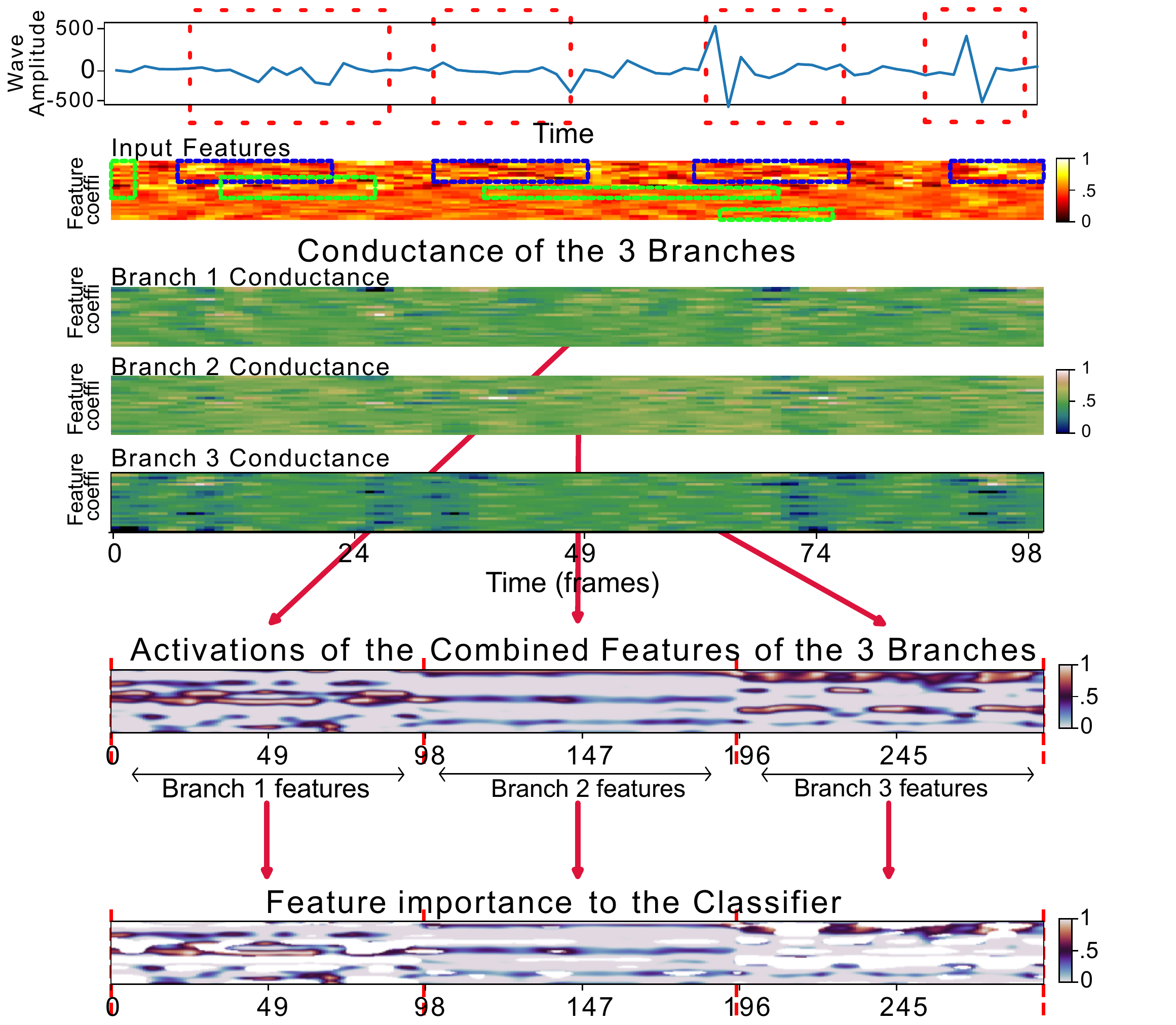}
    \caption{Complete Interpretation generated for a randomly chosen window of the recording  `steth\_20190801\_10\_54\_04.wav' from the test set of the HF Lung V1 dataset.}
    \label{fig:full_interpretation}
\end{figure}

As illustrated in previous subsections, by combining the interpretations generated by DeepLift \cite{shrikumar2017learning}) and Neuron Conductance \cite{dhamdhere2018important} algorithms we can trace back to the time and frequency coefficients within the input feature representation that are salient for the event detection task, which can be used to identify the relevant respective time intervals in the input lung recording. In Fig. \ref{fig:full_interpretation} we provide one such end-to-end interpretation generated for a randomly chosen window of the recording  `steth\_20190801\_10\_54\_04.wav' from the test set of the HF Lung V1 dataset. Note that this example is for the detection of CAS and we have highlighted the salient regions (in green: branch 1; and blue: branch 3) in the input feature based on its saliency to the classifier.  

\subsection{Qualitative Results}

\begin{figure}[htbp]
    \centering
    \subfloat[][]{\includegraphics[width=.95\linewidth]{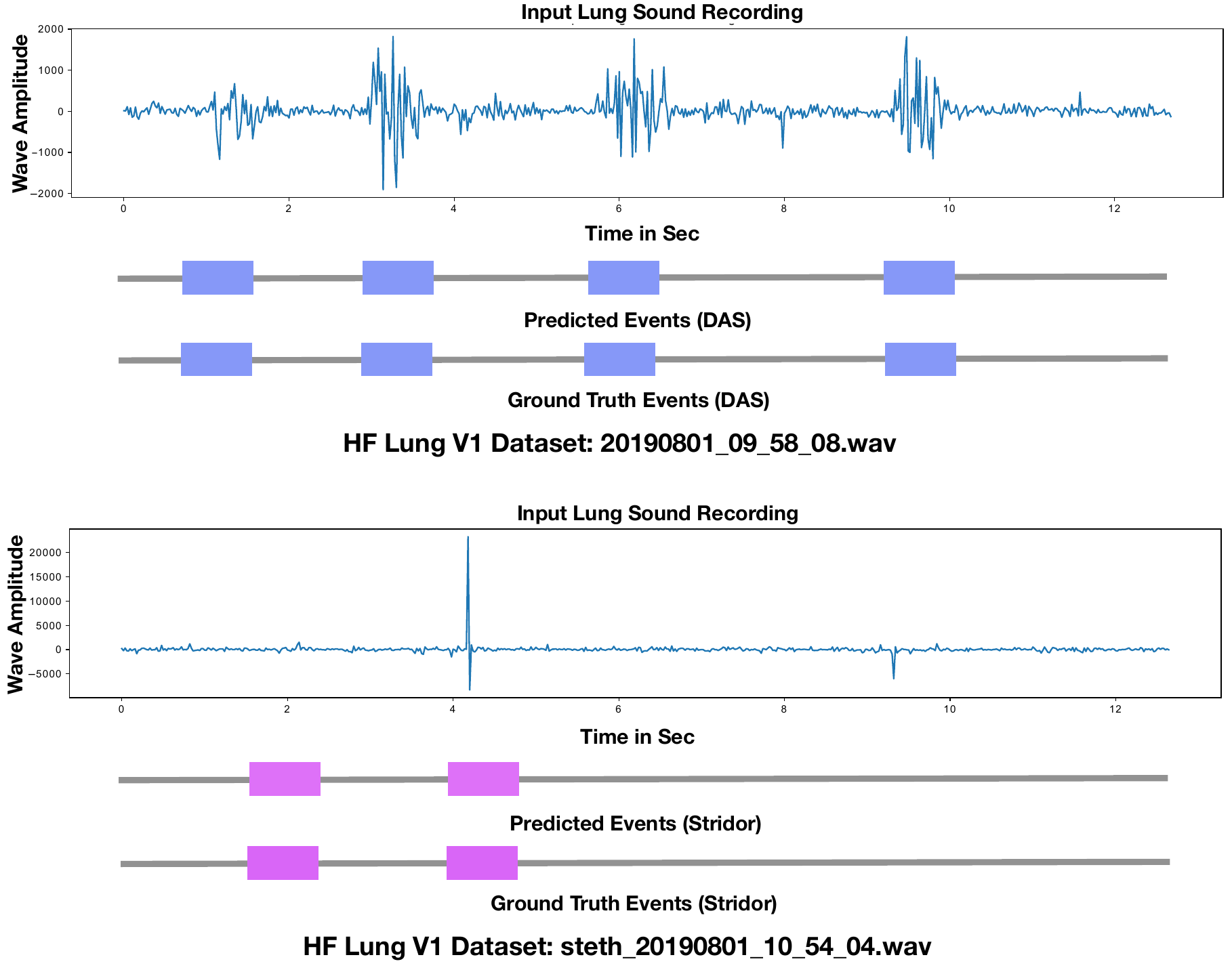}} 
    \caption{Qualitative Results for HF Lung V1 dataset}
    \label{fig:qualitative_1}
\end{figure}

\begin{figure}[htbp]
    \centering
     \subfloat[][]{\includegraphics[width=.95\linewidth]{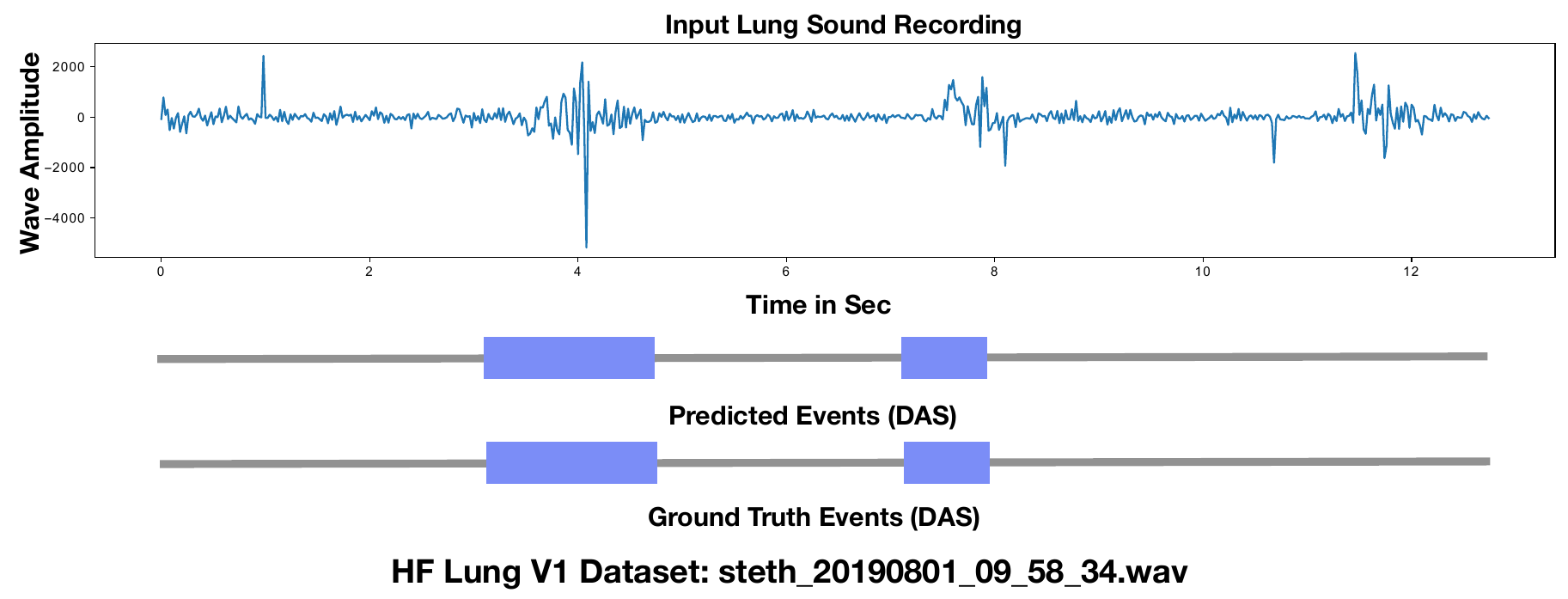}} 
    \caption{Qualitative Results for HF Lung V1 dataset}
    \label{fig:qualitative_2}
\end{figure}

\begin{figure}[htbp]
    \centering
    \subfloat[][]{\includegraphics[width=.95\linewidth]{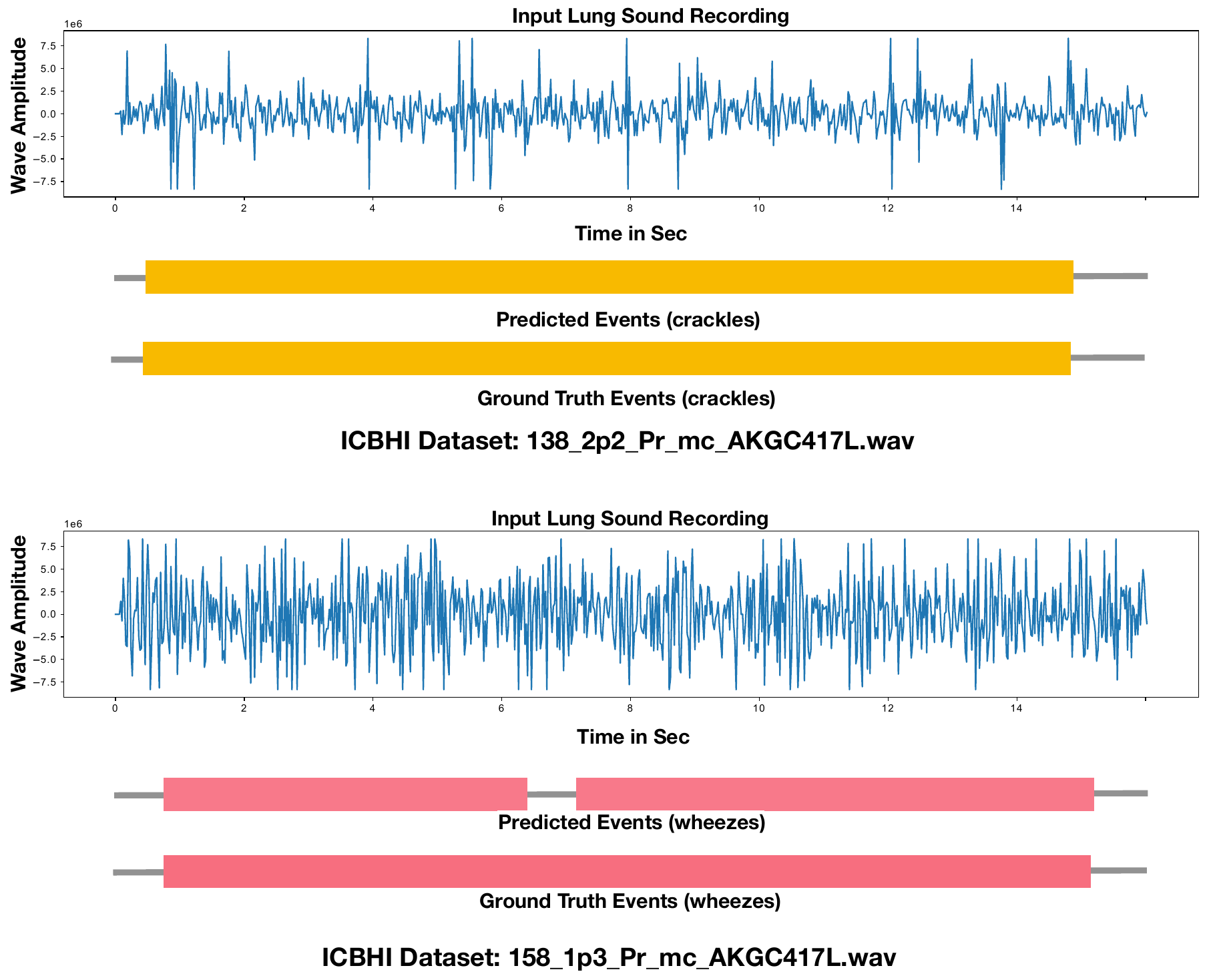}}
    \caption{Qualitative Results for ICBHI dataset}
    \label{fig:qualitative_3}
\end{figure}

\begin{figure}[htbp]
    \centering
    \subfloat[][]{\includegraphics[width=.95\linewidth]{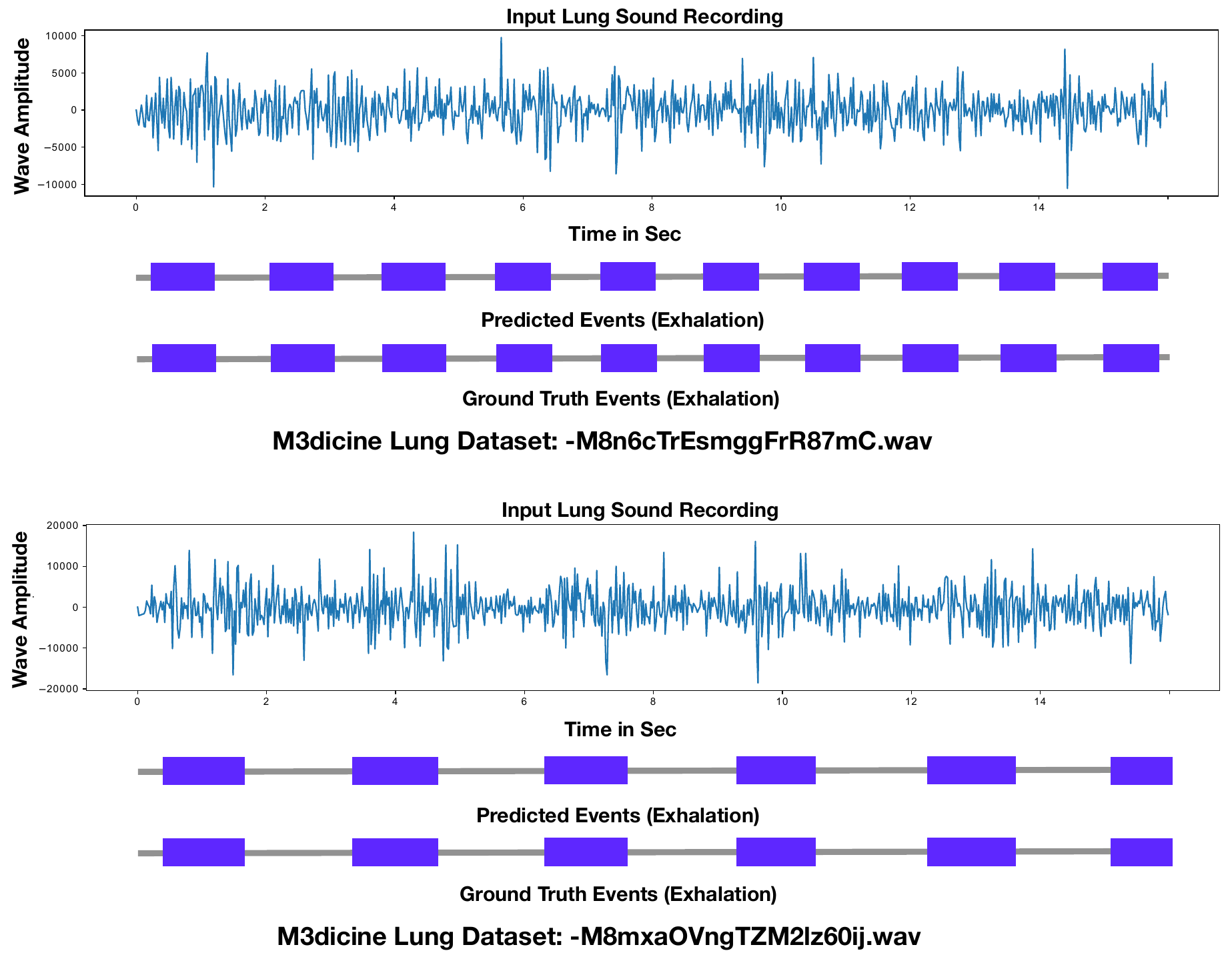}}
    \caption{Qualitative Results for M3dicine Lung dataset}
    \label{fig:qualitative_4}
\end{figure}

In Figs. \ref{fig:qualitative_1}, \ref{fig:qualitative_2}, \ref{fig:qualitative_3}, and \ref{fig:qualitative_4} we show detected events from the proposed model along with the ground truth for randomly selected recordings from the HF Lung V1, ICBHI and M3dicine Lung datasets. Despite the significant difference in the amplitude and noise levels in the audio recordings in different datasets, we observe that the proposed method has been able to generate consistent performance across all considered event categories. We also refer the readers to supplementary material where we provide qualitative results and model interpretations for additional recordings. 

\subsection{Time Complexity}

As indicated earlier the proposed method only contains 276,255 trainable parameters (which is significantly less than the 5,240,513 trainable parameters of the CNN-Bi GRU method of \cite{hsu2021benchmarking}). The proposed framework generates predictions for 100 recordings (each of which has a length of 15 sec) in 1.231054 sec using an Nvidia-GeForce RTX 2080 GPU. Note that this includes time taken for both feature extraction and the generation of model predictions. While these evaluations were conducted using a GPU due to its efficiency, the lightweight nature of our model allows it to be deployed in end user devices such as smart phones and has the ability to generate predictions in real-time.  

\section{Conclusion}
In this paper, we introduced a novel deep learning framework for event detection in respiratory auscultation. We demonstrated how the recent successes in temporal convolution neural network-based temporal modelling, as well as efficient feature strategies, can be incorporated to mitigate the challenges of data scarcity and irregular/noisy recordings. We conducted experiments using multiple benchmarks, including, the HF Lung V1 Database, the ICBHI Respiratory Sound Database and privately collected data (M3DICINE Lung sound database) to evaluate performance for numerous lung sound event detection tasks. The proposed method outperforms the current state-of-the-art methods in all the evaluated event categories, including detection of inhalation, exhalation, crackles, wheeze, stridor, and rhonchi. Moreover, our architecture is simple and efficient in terms of the number of trainable parameters and computational requirements during inference, which affirms its utility in end-user devices such as smartphones for generating real-time predictions. In addition, we provided empirical evaluations of the novel components of the proposed architecture as well as qualitative interpretations of the model prediction, which has been overlooked by existing literature. The evaluations demonstrated the utility of the proposed multi-branch temporal modelling strategy and the effectiveness of the proposed feature fusion mechanism in overcoming the challenges posed by data limitations. Such cost effective implementations, combined with the availability of model interpretations (which increases the trust in the output decisions), make the proposed system suitable for adoption by clinicians for use in hospital settings. Furthermore, we would like to highlight that the proposed framework is not restricted to the analysis of lung sound recordings and could be applied to the analysis of other one-dimensional signals, to detect potential abnormalities in these signals.

\section*{Acknowledgment}
This research was supported by a Cooperative Research Centre Projects (CRC-P) grant. The authors would also like to thank M3DICINE HOLDINGS PTY LTD for providing access to the M3DICINE lung sound database.

\ifCLASSOPTIONcaptionsoff
  \newpage
\fi

\bibliographystyle{IEEEtran}
\bibliography{egdb}

\end{document}